# Diffusivities and Kinetics of Short-Range and Long-Range Orderings in Ni−Fe Permalloys

## T. M. Radchenko, V. A. Tatarenko, and S. M. Bokoch[*,**]

*G. V. Kurdyumov Institute for Metal Physics, N.A.S. of the Ukraine,
36 Academician Vernadsky Blvd.,
UA-03680 Kyyiv-142, Ukraine*
[*]*Taras Shevchenko Kyyiv National University, Physics Department,
2 Academician Glushkov Prospekt,
UA-03022 Kyyiv-22, Ukraine*
[**]*Institut des Materiaux, UFR Sciences et Techniques,
Université de Rouen,
Avenue de l'Université−B.P. 12,
76801 Saint-Etienne du Rouvray Cedex, France*

The microscopic model of atomic diffusion is considered to describe the short-range order relaxation kinetics within the f.c.c.-Ni−Fe Permalloys. The model takes into account both the discrete and anisotropic characters of atomic jumps within the long-range field of concentration heterogeneities of the interacting atoms in an alloy. Experimental data about the time dependence of diffuse-scattering intensity of the thermal neutrons are used to determine the microscopic characteristics of diffusion in alloy at issue. The diffusion coefficients and activation energies for the disordered $Ni_{0.765}Fe_{0.235}$ Permalloy are estimated with the evaluated probabilities of atomic jumps. As shown, the increasing of a Permalloy temperature with a fixed composition influences on the 'potential' field of interatomic interaction (most likely of statistical-force nature) ambiguously: the field 'potential' increases for defined co-ordination shells and decreases for some of other ones. Although the temperature increasing promotes the increasing of any atomic-probabilities jumps generally, but decreasing of the action of 'potential' field generated by the atoms of defined element and caused by its concentration heterogeneities onto the distant sites (co-ordination shells) results in increasing of the atomic-jumps' probabilities of just this element, first of all, into the sites, which are more distant from the 'source' of heterogeneity. Within the framework of the static concentration waves' method along with the self-consistent field approximation, the Onsager-type kinetics equation is obtained to describe the long-range order relaxation by the $L1_2$-type superstructure. To calculate diffusivities for the ordered $Ni_3Fe$ Permalloy, the independent, diffraction experimental data of the long-range order parameter relaxation are used. Theoretical curves of the long-range order time evolution for the non-stoichiometric f.c.c.-$Ni_{3+4(0.25-c_{Fe})}Fe_{1-4(0.25-c_{Fe})}$ Permalloys are plotted. Decreasing of the concentration of alloying element (with a deviation from the stoichiometry) results in decelerating of the long-range order parameter change (on the initial evolution stage) and in increasing of its relaxation time.

Для опису кінетики релаксації близького порядку в пермалоях ГЦК-Ni–Fe розглянуто мікроскопічний модель атомової дифузії. Модель враховує дискретний та анізотропний характери атомових стрибків у далекосяжнім полі концентраційних неоднорідностей взаємодіючих атомів стопу. Для визначення мікроскопічних характеристик дифузії використано експериментальні дані про часову залежність інтенсивності дифузного розсіяння теплових невтронів у стопі. За обчисленими ймовірностями атомових стрибків оцінено коефіцієнти дифузії й енергії активації її у неупорядкованім пермалої $Ni_{0.765}Fe_{0.235}$. Показано, що підвищення температури пермалою фіксованого складу неоднозначно впливає на «потенціяльне» поле міжатомового взаємочину (радше статистично-силової природи): в певних координаційних сферах «потенціял» поля збільшується, а в деяких інших — зменшується. Хоча підвищення температури сприяє збільшенню ймовірности будь-яких стрибків атомів взагалі, але зменшення чину «потенціяльного» поля, що створюється атомами певного первня й обумовлюється його концентраційними неоднорідностями, у вузлах на далеких відстанях (координаційних сферах) призводить до збільшення ймовірности стрибків атомів саме цього первня, насамперед, у більш далекі від «джерела» неоднорідности вузли. В рамках методи статичних концентраційних хвиль і наближення самоузгодженого поля одержано кінетичне рівняння Онсаґерового типу для опису релаксації далекого порядку за надструктурним типом $L1_2$. Задля оцінювання коефіцієнтів дифузії в упорядковнім пермалої $Ni_3Fe$ застосовано незалежні, дифракційні експериментальні дані про зміну у часі параметра далекого порядку. Побудовано теоретичні криві часової еволюції параметра далекого порядку нестехіометричних пермалоїв ГЦК-$Ni_{3+4(0.25-c_{Fe})}Fe_{1-4(0.25-c_{Fe})}$. Зменшення концентрації леґувального первня (з відхиленням від стехіометрії) призводить до уповільнення зміни параметра далекого порядку (на початковій стадії еволюції) й до збільшення часу його релаксації.

Для описания кинетики релаксации ближнего порядка в пермаллоях ГЦК-Ni–Fe рассмотрена микроскопическая модель диффузии атомов. Модель учитывает дискретный и анизотропный характеры атомных прыжков в дальнодействующем поле концентрационных неоднородностей взаимодействующих атомов сплава. Для определения микроскопических характеристик диффузии использованы экспериментальные данные о временной зависимости интенсивности диффузного рассеяния тепловых нейтронов в сплаве. По вычисленным вероятностям атомных прыжков оценены коэффициенты диффузии и энергии активации её в неупорядоченном пермаллое $Ni_{0.765}Fe_{0.235}$. Показано, что повышение температуры пермаллоя фиксированного состава неоднозначно влияет на «потенциальное» поле межатомного взаимодействия (скорее всего статистически-силовой природы): в определенных координационных сферах «потенциал» поля увеличивается, а в некоторых других — уменьшается. Хотя повышение температуры способствует повышению вероятности каких-либо прыжков атомов вообще, но уменьшение действия «потенциально-го» поля, которое образуется атомами определённого элемента и обуслав-ливается его концентрационными неоднородностями, в узлах на далёких расстояниях (координационных сферах) приводит к увеличению вероят-ности прыжков атомов именно этого элемента, прежде всего, в более да-лёкие от «источника» неоднородности узлы. В рамках метода статиче-ских концентрационных волн и приближения самосогласованного поля получено кинетическое уравнение онсагеровского типа для описания ре-лаксации дальнего порядка по сверхструктурному типу $L1_2$. Для оцени-вания коэффициентов диффузии в упорядочивающемся пермаллое $Ni_3Fe$ применены независимые, дифракционные экспериментальные данные об изменении со временем параметра дальнего порядка. Построены теорети-ческие кривые временной эволюции параметра дальнего порядка несте-хиометрических пермаллоев ГЦК-$Ni_{3+4(0,25-c_n)}Fe_{1-4(0,25-c_n)}$. Уменьшение кон-центрации легирующего элемента (с отклонением от стехиометрии) при-водит к замедлению изменения параметра дальнего порядка (на началь-ной стадии эволюции) и к увеличению времени его релаксации.

## 1. INTRODUCTION

Investigation of the Ni–Fe system is important in the wide sense—for understanding of the structure and properties of functional ma-terials as well as the phenomena within the Earth's core. (The Earth's core is surmised to consist mainly of iron–nickel alloy. Therefore, it is of great interest to geophysicists as well.)

The main ordered structure formed in the nickel–iron system is $L1_2$-type superstructure (with a $Ni_3Fe$ stoichiometry) extending for a wide (Permalloy) concentration range. Therefore, we start to study the nickel–iron alloy from the Ni–Fe Permalloys.

Many properties of nickel–iron alloys are determined by their mi-croscopic characteristics and need to be understood at the atomic level. One of such characteristics is a short-range order. It is the unique naturally occurring concentration heterogeneity, whose di-mensions are commensurate with the lattice parameters of an alloy. Kinetics of a short-range order relaxation is determined by the mi-croscopic diffusion over the intersite distances. Therefore, experi-mental measurements of its relaxation kinetics provide us with a possibility to determine the microscopic characteristics of atoms' migration, *i.e.* probabilities and types of atomic jumps.

The goal of a given paper is the extracting of information regard-ing the micro- and macrodiffusivities from the (independent) short-range order and long-range order kinetics' data.

The paper consists of a few sections and subsections. Section 2 is concerned with the short-range order evolution. A model of radiation diffuse-scattering evolution caused by the short-range order relaxation is presented in subsection 2.1. In subsection 2.2, we propose a scheme for the evaluation of both microscopic and macroscopic diffusion characteristics: probabilities of atomic jumps over the f.c.c.-lattice sites and diffusion coefficients. Results obtained within the framework of the proposed scheme are presented in subsection 2.3. 'Potential'-fields' distributions caused by the concentration heterogeneities are determined. Atomic-jumps' probabilities and diffusivities are calculated using the experimental data [1] about the diffuse scattering of thermal neutrons. Section 3 is concerned with the kinetics of a long-range order relaxation. Within the framework of the static-concentration-waves' approach and self-consistent field approximation (subsection 3.1), we use independent diffraction experimental data [2] about the long-range order kinetics to estimate the diffusivities in a $Ni_3Fe$ Permalloy (subsection 3.2). Section 4 contains the discussion and conclusions.

## 2. SHORT-RANGE ORDER RELAXATION

X-ray or thermal-neutron diffraction studies are the most convenient instruments for investigation of the short-range order kinetics [3, 4]. Diffuse scattering of radiations (namely, x-rays or thermal neutrons) is caused by the spatial fluctuations of the atomic-configurations, *i.e.* by the short-range order of atoms. The short-range order relaxation 'promotes' the change of diffuse-scattering intensity with time. Elementary diffusion events can be studied by the time evolution of radiation diffuse scattering intensities. The foreseeable (*in situ*) measurements of short-range discrete diffusion (by means of the short-time elementary diffusion events) can be performed during the realistic time even at the room temperature. Obtained results can be used to determine the low-temperature diffusivities and their activation energies.

### 2.1. Model of Diffuse-Scattering Evolution

If diffusivities of components in a binary $A-B$ alloy are essentially different, one can suppose that the 'fast' atoms of the one kind form a quasi-equilibrium atmosphere around the 'slow' atoms of another kind. Therefore, it is enough to consider the time evolution of the 'slow' $\alpha$-atoms (let $\alpha = B$) [3, 5–8]. This supposition enables to use the $1^{st}$-order kinetics model [3, 5–8] for the time dependence of the diffuse-scattering intensity, $I_{diff}(\mathbf{k}, t)$:

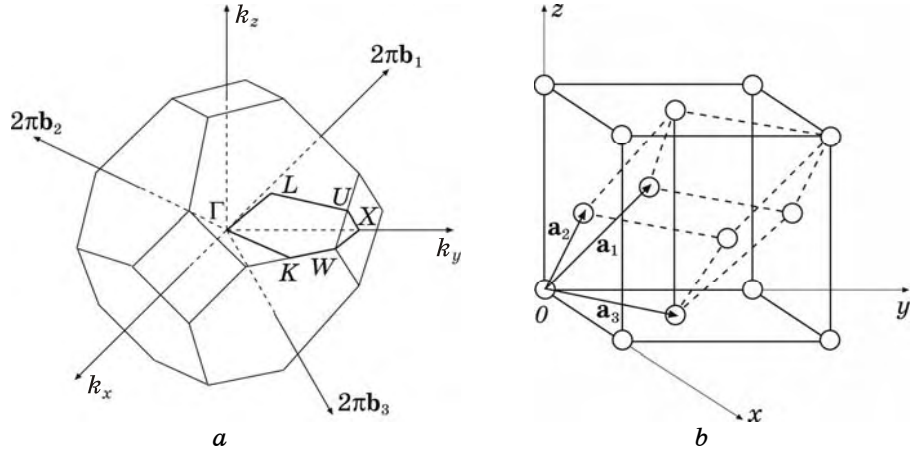

**Fig. 1.** The first Brillouin zone (*a*) ({Γ, *X*, *W*, *L*, *K(U)*}—its high-symmetry points), conditional cubic (large cube) and primitive (interior small parallelepiped) unit cells (*b*) of the Bravais f.c.c. lattice ({$\mathbf{a}_1$, $\mathbf{a}_2$, $\mathbf{a}_3$}—its basis vectors).

$$\frac{I_{\mathrm{diff}}(\mathbf{k},t) - I_{\mathrm{diff}}(\mathbf{k},\infty)}{I_{\mathrm{diff}}(\mathbf{k},0) - I_{\mathrm{diff}}(\mathbf{k},\infty)} \approx \exp\left(-\frac{t}{\tau}\right), \quad \tau \approx \frac{1}{2\lambda_1(\mathbf{k})}. \qquad (1)$$

Here, $I_{\mathrm{diff}}(\mathbf{k},\infty)$ is a diffuse-scattering intensity for the equilibrium alloy; wave vector, $\mathbf{k}$, characterizes the distance of the measurement point in a reciprocal space from the nearest site of reciprocal lattice; $t$ is an instantaneous time, $\tau$ is a relaxation time of the intensity at the $\mathbf{k}$ point (Fig. 1*a*). $\lambda_1(\mathbf{k})$ represents the Fourier transform (with a sign 'minus') of the probabilities of atomic jumps (per unite time) for 'slow' $\alpha$-component into the direct-lattice site with a radius-vector $\mathbf{R} = N_1\mathbf{a}_1 + N_2\mathbf{a}_2 + N_3\mathbf{a}_3$ ($N_1, N_2, N_3$—integers; see Fig. 1*b*).

Thus, the diffuse-scattering kinetics data enable to estimate the relaxation time, $\tau$, and to calculate $\lambda_1(\mathbf{k})$.

## 2.2. Scheme for Determination of Micro- and Macrodiffusivities

If $-\lambda_1(\mathbf{k})$ is known, one can obtain its Fourier originals, $\{-\Lambda_\alpha(\mathbf{R})\}$, which represent the probabilities of the $\alpha$-atoms' jumps per unit time into the site $\mathbf{R}$ from the all surrounding sites $\{\mathbf{R}'\}$ in a field of the interaction 'potential', $\psi_\alpha(\mathbf{R}')$. We assume that the 'potential', $\psi_\alpha(\mathbf{R}')$, at the site $\mathbf{R}'$ is generated, if the microscopic concentration heterogeneities take place, for instance, because of the location of $\alpha$-atom at the 'zero' ('central') site. Thus, $\psi_\alpha(\mathbf{R}')$ represents a 'potential'-field action due to the concentration heterogeneities (of the short-range order type) in a non-ideal alloy. We have the ideal alloy, if there is no such an action.

Apparently, $\Lambda_\alpha(\mathbf{R})$ for any $\mathbf{R}$ (including the 'zero' site) in a crystal lattice is proportional to the all $\{\psi_\alpha(\mathbf{R}')\}$ values. Using inverse Fourier transformation into the $\mathbf{R}$-space, $\Lambda_\alpha(\mathbf{R})$ can be written as follows [9, 10]:

$$\Lambda_\alpha(\mathbf{R}) \equiv \frac{c(1-c)}{k_B T} \sum_{\mathbf{R}'} \Lambda_\alpha^0(\mathbf{R} - \mathbf{R}') \psi_\alpha(\mathbf{R}'), \qquad (2)$$

where $k_B$ is the Boltzmann constant, $T$ is a temperature; $-\Lambda_\alpha^0(\mathbf{R} - \mathbf{R}')$ represents the probability of an $\alpha$-atom jump (per unit time) from any site $\mathbf{R}'$ into the site $\mathbf{R}$ within the ideal solid solution, where the 'potentials' $\{\psi_\alpha(\mathbf{R}')\}$ are equal to $\psi_0 \equiv k_B T/[c(1-c)]$ ($c$ is an atomic fraction of the $\alpha$-kind atoms). Value of $\Lambda_\alpha(\mathbf{R})$ is dependent on a sites' arrangement in a crystal of a given syngony, *i.e.* on a set of the possible differences, $\{\mathbf{R} - \mathbf{R}'\}$, for each $\mathbf{R}$.

In case of the vacancy-diffusion mechanism, we can take into account atomic jumps only over distances between the nearest sites (Fig. 2*a*). When it is necessary to test the possibility of other diffusion mechanism, *i.e.* more complex diffusion processes, we have to consider the several sets of the values: $\{\Lambda_\alpha^0(\mathbf{R} - \mathbf{R}'_I)\}$, $\{\Lambda_\alpha^0(\mathbf{R} - \mathbf{R}'_{II})\}$, *etc*. Indexes I, II, *etc*. relate to the jumps into the site $\mathbf{R}$ from the nearest-neighbour sites $\{\mathbf{R}'_I\}$, next-nearest-neighbour sites $\{\mathbf{R}'_{II}\}$, *etc*.

Analogous models have been considered in [6, 9, 10]. In Ref. [6], only 'algorithm' has been proposed, but diffusivities have been not evaluated. (In Ref. [9], authors have considered the atomic jumps only within the first co-ordination shell in b.c.c. lattice.)

Let us make the following assumptions. Firstly, $\psi_\alpha(\mathbf{R}')$ is a non-vanishing function only within the six co-ordination shells around the 'zero' site. Secondly, probability of atomic jump is a non-zero value only for the two co-ordination shells. Thus,

$$\Lambda_\alpha^0(\mathbf{0}) = \Lambda_{\alpha 0}^0 \neq 0, \quad \Lambda_\alpha^0(R_I) = \Lambda_{\alpha I}^0 \neq 0, \quad \Lambda_\alpha(R_{II}) = \Lambda_{\alpha II}^0 \neq 0,$$

$$\psi_\alpha(\mathbf{0}) = \psi_{\alpha 0} \neq 0, \quad \psi_\alpha(R_I) = \psi_{\alpha I} \neq 0, \quad \psi_\alpha(R_{II}) = \psi_{\alpha II} \neq 0,$$

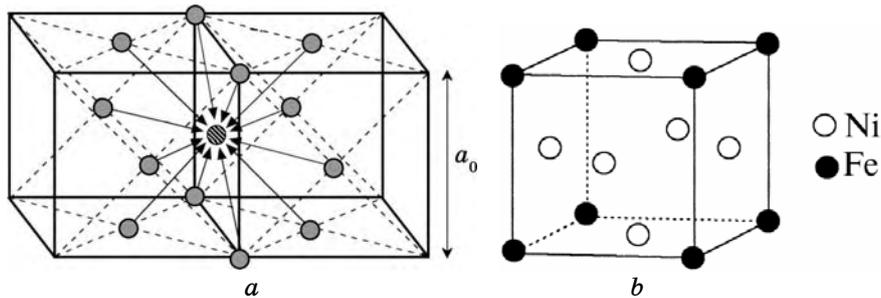

**Fig. 2.** Atomic jumps into the given site (a striped ball) from the nearest sites (gray balls) in f.c.c. lattice (*a*) and $L1_2$-type superstructure of $Ni_3Fe$ (*b*).

$$\psi_\alpha(R_{III}) = \psi_{\alpha III} \neq 0, \quad \psi_\alpha(R_{IV}) = \psi_{\alpha IV} \neq 0,$$

$$\psi_\alpha(R_V) = \psi_{\alpha V} \neq 0, \quad \psi_\alpha(R_{VI}) = \psi_{\alpha VI} \neq 0,$$

where $R_I$, $R_{II}$, *etc.* are the radii of the 1$^{st}$, 2$^{nd}$, *etc.* co-ordination shells, respectively (the other probabilities, $\Lambda^0_{\alpha III}$, $\Lambda^0_{\alpha IV}$, *etc.*, and 'potential' functions, $\psi_{\alpha VII}$, $\psi_{\alpha VIII}$, *etc.*, are equal to zero) For a distinctness, let us suppose that $\alpha$-atom is located at the 'zero' site (with $\mathbf{R} = \mathbf{0}$). Then, one can write as follows:

$$\Lambda_\alpha\left(\mathbf{R}_0(000)\right) \cong 12\Lambda^0_{\alpha I}\left[\frac{\psi_{\alpha I}}{\psi_0}\right] + 6\Lambda^0_{\alpha II}\left[\frac{\psi_{\alpha II}}{\psi_0}\right] + \Lambda^0_{\alpha 0}\left[\frac{\psi_{\alpha 0}}{\psi_0}\right],$$

$$\Lambda_\alpha\left(\mathbf{R}_I(110)\right) \cong 4\Lambda^0_{\alpha I}\left[\frac{\psi_{\alpha I}}{\psi_0}\right] + 2\Lambda^0_{\alpha I}\left[\frac{\psi_{\alpha II}}{\psi_0}\right] + \Lambda^0_{\alpha I}\left[\frac{\psi_{\alpha III}}{\psi_0}\right] +$$

$$+4\Lambda^0_{\alpha I}\left[\frac{\psi_{\alpha IV}}{\psi_0}\right] + 2\Lambda^0_{\alpha II}\left[\frac{\psi_{\alpha I}}{\psi_0}\right] + \Lambda^0_{\alpha 0}\left[\frac{\psi_{\alpha I}}{\psi_0}\right] + \Lambda^0_{\alpha I}\left[\frac{\psi_{\alpha 0}}{\psi_0}\right],$$

$$\Lambda_\alpha\left(\mathbf{R}_{II}(200)\right) \cong 4\Lambda^0_{\alpha I}\left[\frac{\psi_{\alpha I}}{\psi_0}\right] + 4\Lambda^0_{\alpha I}\left[\frac{\psi_{\alpha IV}}{\psi_0}\right] + \Lambda^0_{\alpha II}\left[\frac{\psi_{\alpha VI}}{\psi_0}\right] +$$

$$+4\Lambda^0_{\alpha II}\left[\frac{\psi_{\alpha III}}{\psi_0}\right] + \Lambda^0_{\alpha 0}\left[\frac{\psi_{\alpha II}}{\psi_0}\right] + \Lambda^0_{\alpha II}\left[\frac{\psi_{\alpha 0}}{\psi_0}\right],$$

$$\Lambda_\alpha\left(\mathbf{R}_{III}(211)\right) \cong 2\Lambda^0_{\alpha I}\left[\frac{\psi_{\alpha I}}{\psi_0}\right] + \Lambda^0_{\alpha I}\left[\frac{\psi_{\alpha II}}{\psi_0}\right] + 2\Lambda^0_{\alpha I}\left[\frac{\psi_{\alpha III}}{\psi_0}\right] + 2\Lambda^0_{\alpha I}\left[\frac{\psi_{\alpha IV}}{\psi_0}\right] +$$

$$+\Lambda^0_{\alpha I}\left[\frac{\psi_{\alpha V}}{\psi_0}\right] + \Lambda^0_{\alpha II}\left[\frac{\psi_{\alpha I}}{\psi_0}\right] + 2\Lambda^0_{\alpha II}\left[\frac{\psi_{\alpha IV}}{\psi_0}\right] + \Lambda^0_{\alpha 0}\left[\frac{\psi_{\alpha IV}}{\psi_0}\right],$$

$$\Lambda_\alpha\left(\mathbf{R}_{IV}(220)\right) \cong \Lambda^0_{\alpha I}\left[\frac{\psi_{\alpha I}}{\psi_0}\right] + 4\Lambda^0_{\alpha I}\left[\frac{\psi_{\alpha IV}}{\psi_0}\right] + 2\Lambda^0_{\alpha II}\left[\frac{\psi_{\alpha II}}{\psi_0}\right] +$$

$$+2\Lambda^0_{\alpha II}\left[\frac{\psi_{\alpha V}}{\psi_0}\right] + \Lambda^0_{\alpha 0}\left[\frac{\psi_{\alpha III}}{\psi_0}\right],$$

$$\Lambda_\alpha\left(\mathbf{R}_V(310)\right) \cong \Lambda^0_{\alpha I}\left[\frac{\psi_{\alpha II}}{\psi_0}\right] + \Lambda^0_{\alpha I}\left[\frac{\psi_{\alpha III}}{\psi_0}\right] + \Lambda^0_{\alpha I}\left[\frac{\psi_{\alpha IV}}{\psi_0}\right] +$$

$$+\Lambda^0_{\alpha I}\left[\frac{\psi_{\alpha VI}}{\psi_0}\right] + \Lambda^0_{\alpha II}\left[\frac{\psi_{\alpha I}}{\psi_0}\right],$$

$$\Lambda_\alpha \left( \mathbf{R}_{\mathrm{VI}}(222) \right) \cong 3\Lambda_{\alpha\mathrm{I}}^0 \left[ \frac{\psi_{\alpha\mathrm{IV}}}{\psi_0} \right] + 3\Lambda_{\alpha\mathrm{II}}^0 \left[ \frac{\psi_{\alpha\mathrm{III}}}{\psi_0} \right] + \Lambda_{\alpha0}^0 \left[ \frac{\psi_{\alpha\mathrm{V}}}{\psi_0} \right],$$

$$\Lambda_\alpha \left( \mathbf{R}_{\mathrm{VII}}(321) \right) \cong \Lambda_{\alpha\mathrm{I}}^0 \left[ \frac{\psi_{\alpha\mathrm{III}}}{\psi_0} \right] + \Lambda_{\alpha\mathrm{I}}^0 \left[ \frac{\psi_{\alpha\mathrm{IV}}}{\psi_0} \right] + \Lambda_{\alpha\mathrm{I}}^0 \left[ \frac{\psi_{\alpha\mathrm{V}}}{\psi_0} \right] + \Lambda_{\alpha\mathrm{II}}^0 \left[ \frac{\psi_{\alpha\mathrm{IV}}}{\psi_0} \right],$$

$$\Lambda_\alpha \left( \mathbf{R}_{\mathrm{VIII}}(400) \right) \cong 4\Lambda_{\alpha\mathrm{I}}^0 \left[ \frac{\psi_{\alpha\mathrm{VI}}}{\psi_0} \right] + \Lambda_{\alpha\mathrm{II}}^0 \left[ \frac{\psi_{\alpha\mathrm{II}}}{\psi_0} \right].$$

In these expressions for $\Lambda_\alpha(\mathbf{R}_\mathrm{n}(lmn))$, $(lmn)$ are the co-ordinates of the sites in the usual cubic-lattice system with the [100], [010], [001] translation vectors along the directions of $0x$, $0y$, $0z$ axes, respectively (Fig. 1$b$); co-ordinates $(lmn)$ are understood in the units of $a_0/2$ where $a_0$ is an f.c.c.-lattice parameter (Fig. 2). $-\Lambda_{\alpha0}^0$ is a probability (per unite time) for the $\alpha$-atom to stay in a site at issue; $\psi_{\alpha0}$ is a 'potential' function value at the 'zero' site.

The values of $\Lambda_\alpha(\mathbf{R}_\mathrm{n}(lmn))$ can be evaluated from the inverse Fourier transform by the values of $\lambda_1(\mathbf{k})$ estimated on the basis of the kinetics model (1) using the experimental data for the disordered f.c.c.-$^{62}$Ni$_{0.765}$Fe$_{0.235}$ Permalloy [1].

The Fourier original of a probability for atom to jump into the site $\mathbf{R}$ of an f.c.c. lattice is as follows:

$$\Lambda_\alpha(\mathbf{R}) = K(lmn) \sum_{k_1 k_2 k_3} \lambda_1(k_1 k_2 k_3) \cos(2\pi k_1 l) \cos(2\pi k_2 m) \cos(2\pi k_3 n),$$

where $K$ is a geometrical coefficient dependent on the $\mathbf{R}(lmn)$.

The atomic-jumps' probabilities allow to calculate the macroscopic diffusion characteristics, *i.e.* the diffusion and self-diffusion coefficients of the 'slow' $\alpha$-atoms. Both the long-wave limit transition from the discrete migration process to the continual transfer and the assumption of the equiprobable atomic jumps into the sites within the same co-ordination shell (relative to the 'zero' site) yield formulas for the diffusivities in ideal ($D_\alpha^*$) and non-ideal ($D_\alpha$) cubic solutions [5–9]:

$$D_\alpha^* \approx -\frac{1}{6} \sum_{\mathrm{n=I}}^\infty \Lambda_\alpha^0(R_\mathrm{n}) R_\mathrm{n}^2 Z_\mathrm{n}, \quad D_\alpha \approx -\frac{1}{6} \sum_{\mathrm{n=I}}^\infty \Lambda_\alpha(R_\mathrm{n}) R_\mathrm{n}^2 Z_\mathrm{n}, \tag{3}$$

where $Z_\mathrm{n}$ is a co-ordination number for the n-th co-ordination shell.

### 2.3. Results of Simulation

Above-presented scheme can be used to calculate the diffusion characteristics of Ni$_{0.765}$Fe$_{0.235}$-type Permalloy, which is similar (by the composition) to the $L1_2$-Ni$_3$Fe considered hereinafter (Fig. 2$b$).

In Figure 3, the jumps' probabilities for the 'slow' α-atoms (*i.e.* conditionally 'slow' Fe atoms within the 'coat' of 'fast' Ni atoms [11]) in the ideal and non-ideal $Ni_{0.765}Fe_{0.235}$ Permalloys are shown at the temperatures of 776 K and 783 K.

The first column in Fig. 3*a* means a probability for 'slow' atom to remain (during a time unite) at a given site. One can readily see that magnitude of this probability ($\approx 0.024$ s$^{-1}$ and $0.033$ s$^{-1}$ for 776 K and 783 K, respectively) as well as of two other probabilities for the first and second co-ordination shells is smaller than for f.c.c.-Ni–Mo alloy ($\approx 0.6$ s$^{-1}$ [10]).

The atomic jumps in the non-ideal Permalloy (Fig. 3*b*) are deter-

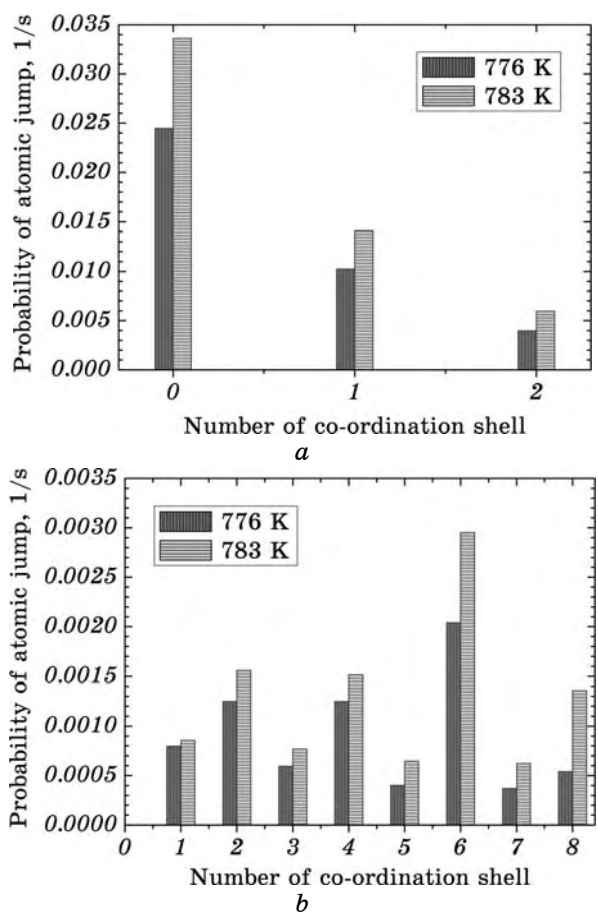

*a*

*b*

**Fig. 3.** Probabilities of atomic jumps (per unite time) into the different co-ordination shells (ambient a 'zero' site) for the ideal (*a*) and non-ideal (*b*) disordered $^{62}Ni_{0.765}Fe_{0.235}$ Permalloy. The heights of the columns indicate the magnitudes of $-\Lambda_{\alpha}^{0}(R_{n})$ (*a*) and $-\Lambda_{\alpha}(R_{n})$ (*b*) mentioned in the text.

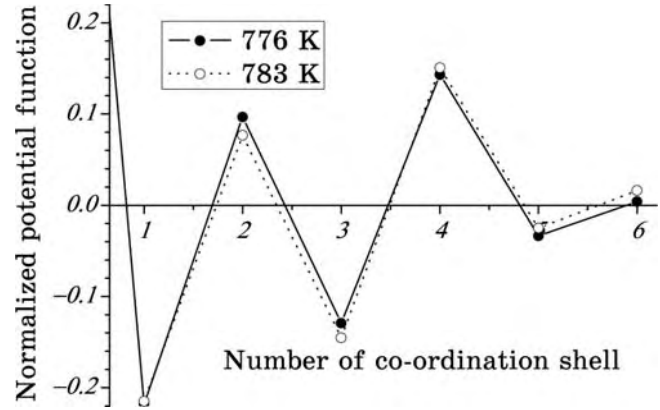

**Fig. 4.** Values of the normalized 'potential' functions, $\psi_\alpha(R_n)/\psi_0$, at the different co-ordination shells (ambient a 'zero' site) in $^{62}\text{Ni}_{0.765}\text{Fe}_{0.235}$.

**TABLE 1.** Vacancy-controlled diffusion ($D_{\text{Fe}}$), self-diffusion ($D_{\text{Fe}}^*$), and interdiffusion ($D$) coefficients for $^{62}\text{Ni}_{0.765}\text{Fe}_{0.235}$ Permalloy.

| $T$, K | $D_{\text{Fe}}$, cm$^2$/s | $D_{\text{Fe}}^*$, cm$^2$/s | $D$, cm$^2$/s [3] |
|--------|---------------------------|------------------------------|-------------------|
| 776 | $4.49 \cdot 10^{-17}$ | $1.81 \cdot 10^{-17}$ | $2.49 \cdot 10^{-18}$ |
| 783 | $6.90 \cdot 10^{-17}$ | $2.55 \cdot 10^{-17}$ | $3.56 \cdot 10^{-18}$ |

mined by the 'potential' functions (Fig. 4) caused by the concentration heterogeneities. It has quasi-oscillating character, and its absolute value decreases overall with an increasing distance from the concentration heterogeneity at the 'zero' site (Fig. 4). At the 'zero' site (for the 'zero' co-ordination shell), $\psi_\alpha(\mathbf{0})/\psi_0 = \psi_{00}/\psi_0 \approx 1$.

The diffusion and self-diffusion coefficients of Fe atoms in f.c.c.-$^{62}\text{Ni}_{0.765}\text{Fe}_{0.235}$ calculated from Eq. (3) are listed in Table 1. In the last column, the interdiffusion coefficients extrapolated from the high-temperature measurements [1] are presented as well. The total activation energies of the vacancy-mediated diffusion and self-diffusion of Fe atoms are calculated too as 3.21 eV and 2.56 eV, respectively.

## 3. LONG-RANGE ORDER EVOLUTION

Let us now consider the case of an exchange ('ring') mechanism [3, 8, 14–20] governing the diffusion relaxation of $L1_2$-type ordering Ni$_3$Fe Permalloy [2] (at the temperatures below the temperature of the order–disorder phase transformation). This Permalloy is similar by the composition to the above-considered Ni$_{0.765}$Fe$_{0.235}$.

### 3.1. Model

The atomic distribution function in this ordering $A_{1-c}B_c$ alloy can be represented as a superposition of the static concentration waves [3].

Using the static concentration waves' approach along with the self-consistent-field approximation [3] and the Onsager-type kinetics equation for a diffusion relaxation of distribution of the atoms of both $A$ and $B$ over the crystal lattice sites, the differential kinetics equation for relaxation of long-range order parameter, $\eta(t, T, c)$, with time, $t$, for the $L1_2$-type structure is as follows:

$$\frac{d\eta}{dt} = -c(1-c)\tilde{L}_{\circ}(\mathbf{k}_X)\left(\frac{\tilde{w}(\mathbf{k}_X)}{k_B T}\eta + \ln\frac{\left(c + \dfrac{3}{4}\eta\right)\left(1 - c + \dfrac{1}{4}\eta\right)}{\left(1 - c - \dfrac{3}{4}\eta\right)\left(c - \dfrac{1}{4}\eta\right)}\right). \qquad (4)$$

Here, $\mathbf{k}_X = (100)$ is a wave vector, which generates the $L1_2$-type superstructure and which belongs to the first Brillouin zone for the f.c.c. lattice (Fig. 1); $\tilde{L}_{\circ}(\mathbf{k}_X)$ is the corresponding Fourier transform of the Onsager 'ring'-kinetics coefficients, $\tilde{w}(\mathbf{k}_X)$ is the Fourier transform of the pairwise-interaction mixing (interchange) energies. Magnitudes of $\tilde{w}(\mathbf{k}_X)$ for the $Ni_3Fe$ Permalloy at different temperatures were estimated in [6].

An implicit solution $t = t(\eta)$ of Eq. (4) is given by the following expression:

$$t = -\frac{1}{c(1-c)\tilde{L}_{\circ}(\mathbf{k}_X)}\int_{\eta_0}^{\eta}\left(\frac{\tilde{w}(\mathbf{k}_X)}{k_B T}\eta' + \ln\frac{\left(c + \dfrac{3}{4}\eta'\right)\left(1 - c + \dfrac{1}{4}\eta'\right)}{\left(1 - c - \dfrac{3}{4}\eta'\right)\left(c - \dfrac{1}{4}\eta'\right)}\right)^{-1} d\eta', \quad (5)$$

where $\eta_0$ is an initial (non-equilibrium) magnitude of the long-range order parameter (at $t = 0$, when relaxation starts after the alloy quenching).

Assuming the atomic jumps between only the nearest-neighbour sites in f.c.c. lattice and using the condition of a fixed total number of the atoms in a system, for kinetics coefficients, we can write:

$$\tilde{L}_{\circ}(\mathbf{k}) \approx -4L_{\circ}(R_1)[3 - \cos(\pi k_1)\cos(\pi k_2) -$$

$$-\cos(\pi k_2)\cos(\pi k_3) - \cos(\pi k_3)\cos(\pi k_1)]; \qquad (6)$$

here, $-L_{\circ}(R_1)$ is proportional to the 'ring'-jumps' probability for a pair of the atoms at the nearest-neighbour sites, $\mathbf{R}$ and $\mathbf{R}'$, per unite time $(R_1 = |\mathbf{R} - \mathbf{R}'|)$.

## 3.2. Results of Computation

Using experimental data [2] (Fig. 5) and Eq. (5), we estimated the time-optimized values of $L_\circ(\mathbf{k}_X)$ at 673 K and 743 K (see Table 2). Then, values of $-L_\circ(R_I)$ were calculated from expression (6). Their time-optimized values are the following: $-L_\circ(R_I) \approx 4.27 \cdot 10^{-7}\,\mathrm{s}^{-1}$ and $-L_\circ(R_I) \approx$ $\approx 5.785 \cdot 10^{-6}\,\mathrm{s}^{-1}$ at $T = 673$ K and $T = 743$ K, respectively. Substituting values of $-L_\circ(R_n)$ (with $-L_\circ(R_n) \equiv 0$ for $n \geq \mathrm{II}$) instead of $-\Lambda_\alpha(R_n)$ into equation similar to Eq. (3), we estimated roughly the exchange diffusion mobilities of Fe and Ni atoms in their pairs within the $L1_2$-type Ni$_3$Fe ($D_\circ = 1.084 \cdot 10^{-21}\,\mathrm{cm}^2/\mathrm{s}$ and $D_\circ = 1.469 \cdot 10^{-20}\,\mathrm{cm}^2/\mathrm{s}$ for $T = 673$ K and $T = 743$ K, respectively). Evaluated diffusion-migration activation energy for Fe (and Ni) atoms is 1.60 eV (according to the Arrhenius formula).

Annealing of the Ni$_3$Fe alloy at 673 K and 743 K results in an increase in size of short-range ordered 'domains' (clusters). Their sizes at 673 K increase comparatively slow and run up to 3.5 nm, while at 743 K, they run up to 20 nm during the annealing of $3.6 \cdot 10^5$ s in both cases [2]. The short-range ordered 'domains' growth is a more long-standing process than short-range ordering within the domains. However, in substitutional alloys, the long-range order is formed by means of accretion of the short-range order of substitution [3, 4]. Therefore, the slowness of the $\eta$ increasing at 673 K, apparently, is conditioned mainly by the rate of the short-range ordered domains' overgrowth. Within the framework of the one-exponent kinetics model (see solid and dashed curves in Fig. 5), relaxation time is $\approx 2.5 \cdot 10^4$ s at 743 K, while at 673 K, it is $\approx 3.9 \cdot 10^4$ s. After the $3.6 \cdot 10^5$ s of annealing, $\eta \approx$

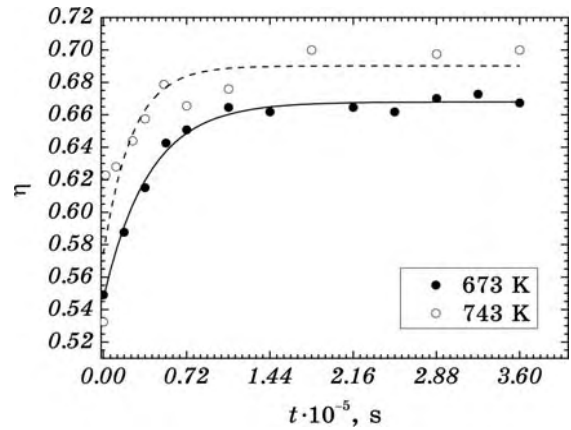

**Fig. 5.** Long-range order parameter, $\eta$, *vs.* time, $t$, for Ni$_3$Fe Permalloy: open and full circles represent experimental points [2], dashed and solid curves—their exponential smoothing.

**TABLE 2.** Onsager exchange-kinetics coefficients (optimized according to all available experimental data about $S_l$* [2] for different annealing times $t_l$; $l = 1$, 2, …) for the $L1_2$-type Ni$_3$Fe alloy at the two annealing temperatures.

| $t_l \cdot 10^{-5}$, s | $T = 673$ K | $T = 743$ K |
|---|---|---|
| | $S_l \cdot 10^6$, s$^{-1}$ | $S_l \cdot 10^6$, s$^{-1}$ |
| 0.02 | | 615.0 |
| 0.11 | | 108.9 |
| 0.18 | 14.53 | |
| 0.25 | | 55.09 |
| 0.36 | 12.41 | 43.68 |
| 0.52 | | 36.15 |
| 0.54 | 11.72 | |
| 0.72 | 9.572 | 23.43 |
| 1.08 | 7.259 | 17.09 |
| 1.44 | 5.312 | |
| 1.80 | | 12.42 |
| 2.16 | 3.630 | |
| 2.52 | 3.035 | |
| 2.88 | 2.855 | 7.605 |
| 3.24 | 2.597 | |
| 3.60 | 2.231 | 6.212 |
| | $\tilde{L}_\circ(\mathbf{k}_X) \equiv \dfrac{1}{11}\sum\limits_{l=1}^{11} S_l = 6.832\cdot 10^{-6}\,\text{s}^{-1}$ | $\tilde{L}_\circ(\mathbf{k}_X) \equiv \dfrac{1}{10}\sum\limits_{l=1}^{10} S_l = 9.256\cdot 10^{-5}\,\text{s}^{-1}$ |

$$* \; S_l \equiv -\frac{16}{3t_l}\int\limits_{\eta_0}^{\eta_l}\left(\frac{\tilde{w}(\mathbf{k}_X)}{k_B T}\eta' + \ln\frac{(1+3\eta')(3+\eta')}{3(1-\eta')^2}\right)^{-1} d\eta' \quad (c = 1/4).$$

$\approx 0.70$ at 743 K, while $\eta \approx 0.67$ at 673 K (Fig. 5) [2]. At least, it means that an *equilibrium* value of the long-range order parameter was *not* reached in experiment [2] at 673 K.

Figures 6*a*, 6*b* demonstrate the last statement. In these figures, theoretical dependences of the long-range order parameter on time are obtained by the numerical solution of Eq. (4) with the optimized $\tilde{L}_\circ(\mathbf{k}_X)$ (see Table 2). As shown in Figs. 6*a*, 6*b*, for small annealing times, instantaneous (*non-equilibrium*) magnitude of long-range order parameter at 673 K is lower than it is for 743 K. Nevertheless, a long-time annealing leads to the higher (near-equilibrium) long-range order parameter at 673 K than it is at 743 K (Figs. 6).

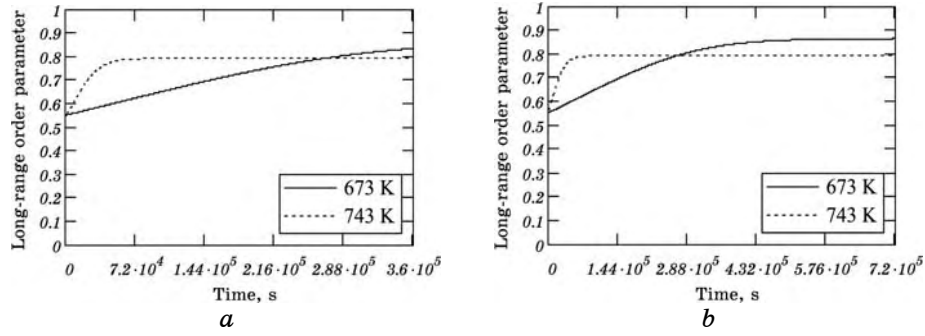

**Fig. 6.** Theoretical dependence of long-range order parameter, $\eta$, on time, $t$, (in duplicate of scales) for Ni$_3$Fe Permalloy at two temperatures.

Equation (4) can be solved numerically for a wide temperature–concentration range. The curves in Figs. 7, 8 represent the solutions of Eq. (4) at the different reduced temperatures, $T^* = k_B T / |\hat{w}(\mathbf{k}_X)|$, and at the two (real) absolute temperatures, $T$, respectively. (Dependence of a long-range order parameter on the $T$-dependent reduced-time, $t^* = \bar{L}_{\circ}(\mathbf{k}_X)t$, is shown in both Figs. 7, 8.)

We also solved Eq. (4) using both the optimized values of $\bar{L}_{\circ}(\mathbf{k}_X)$ and the calculated magnitudes of $\hat{w}(\mathbf{k}_X)$ [6, 13] to forecast the evolution of $\eta$ (in real time, $t$) for the non-stoichiometric Ni$_{1-c}$Fe$_c$ Permalloys ($c < 1/4$). The resulting values of $\eta(t)$ are plotted as a function of $t$ for six initial conditions: $\eta(0) \equiv \eta_0 = 0.1$, $\eta_0 = 0.2$, $\eta_0 = 0.3$, $\eta_0 = 0.4$, $\eta_0 = 0.5$ and $\eta_0 = 0.6$. Kinetics curves are shown in Figs. 9, 10. The calculations were not extended for more high initial values of long-range order parameter, since an initial value of $\eta_0$ cannot exceed the equilibrium value of $\eta_{eq}(T, c)$ in case of the ordering. The last assertion allows estimating the maximal admissible initial value of long-range order parameter, $\eta_0^{max}$, which does not exceed the equilibrium one, $\eta_{eq}$. Values of $\eta_{eq}$ ($\geq \eta_0^{max}$) are plotted in Fig. 11 for different atomic fractions, $c = c_{Fe}$, of Fe in a Ni–Fe Permalloy at issue.

### 3.3. Comparison with Statistical-Thermodynamics Model Results

A value of the equilibrium long-range order parameter 'jump' in Fig. 11 may be verified within the framework of the statistical-thermodynamics model [3, 7, 12, 13] by solving the set of equations for the equilibrium conditions of phase transformation:

$$\left.\frac{\partial F}{\partial \eta}\right|_{\substack{T=T_K(c) \\ \eta = \Delta\eta_{eq}}} = 0, \quad F(\Delta\eta_{eq}, T_K) = F(0, T_K);$$

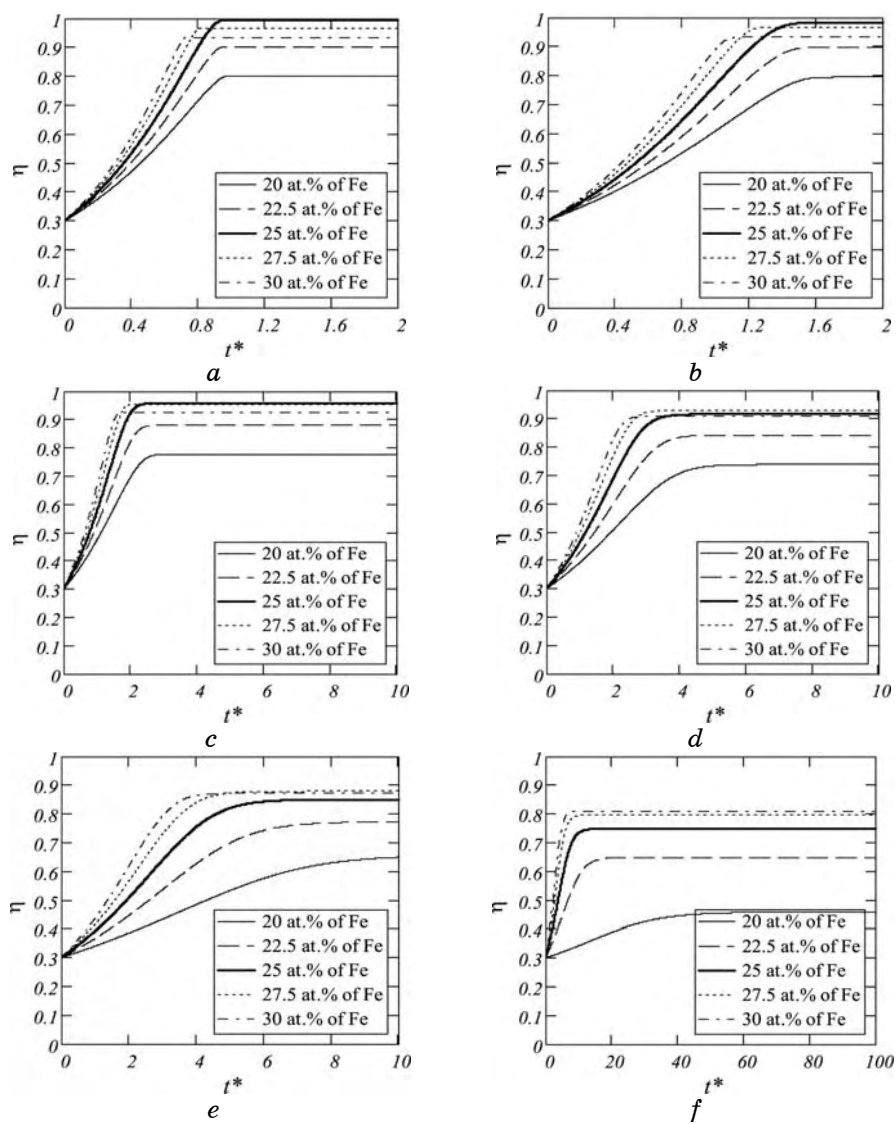

**Fig. 7.** Long-range order parameter, η, *vs.* the reduced annealing time, *t**, for Ni−Fe Permalloys at the different reduced temperatures, *T**: 0.08 (*a*), 0.10 (*b*), 0.12 (*c*), 0.14 (*d*), 0.16 (*e*), and 0.18 (*f*).

here, $F$ is a total configuration-dependent free energy; $F(\Delta\eta_{eq}, T_K)$ and $F(0, T_K)$ are the configurational free energies of alloy in ordered and disordered states, respectively, at the order−disorder phase transformation (Kurnakov's) temperature, $T_K$, when the equilibrium long-range order parameter 'jump', $\Delta\eta_{eq}$, occurs.

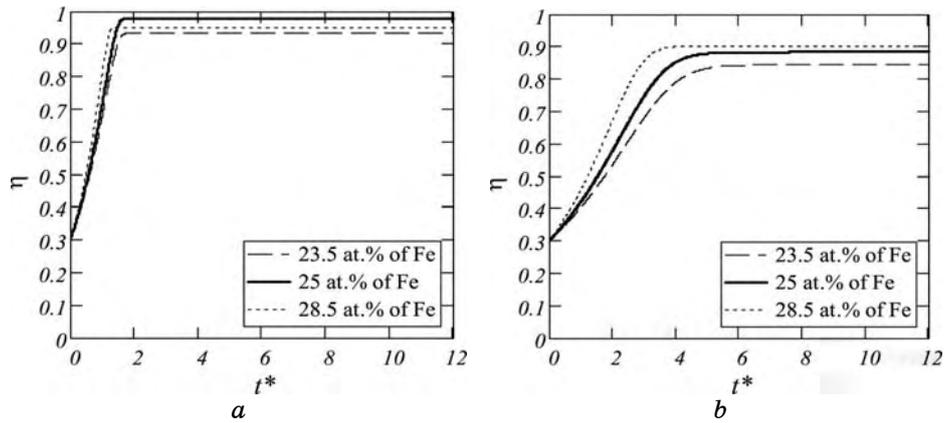

**Fig. 8.** Theoretical dependence of a long-range order parameter, η, on the reduced time, $t^*$, for the stoichiometric (solid bold curve) and non-stoichiometric (dotted and dashed curves) Ni–Fe Permalloys at the temperatures of 450 K (*a*) and 650 K (*b*).

Numerical computation of the above-mentioned set of equations was made in [7, 12, 13]. There were obtained $c_{Fe}$-dependences of both $\Delta\eta_{eq}$ and reduced Kurnakov's temperature, $T_K^* = k_B T_K / |\dot{w}(\mathbf{k}_X)|$. Those results [7, 12, 13] and our complementary calculations confirm the salient features in Figs. 7, 8, 11.

## 4. DISCUSSION AND CONCLUSIONS

**i)** For the ideal f.c.c.-Ni$_{0.765}$Fe$_{0.235}$-type alloy, probability of atomic jumps into a given site **R** from the nearest sites (for site **R**) is less than $-\Lambda_\alpha(\mathbf{0})/2$, and from the next-nearest sites, it is essentially less than $-\Lambda_\alpha(\mathbf{0})/2$ (Fig. 3*a*). This means the predominance of the atomic jumps within the first co-ordination shell and possibility of diffusion mainly governed by the vacancy mechanism within the commonly accepted interpretation.

**ii)** For the non-ideal f.c.c.-Ni$_{0.765}$Fe$_{0.235}$-type Permalloy, probability of the 'slow' α-atom jump into the site **R**, $-\Lambda_\alpha(\mathbf{R})$, is determined by the field at the site **R**. That is why the probability of jumps of the α-atoms into the site **R** is non-monotone function of **R** (Fig. 3*b*): it is higher at sites, where arrangement of the α-atoms is more energy-wise favourable.

Suppose the short-range order of the $L1_2$(Ni$_3$Fe)-type with an f.c.c.-lattice where the 'zero' site coincides with one of the cube corners of the unite cell. If we put Fe-atom at such a cube corner, then the Fe-atoms will try to occupy predominantly the cube corners (*viz.* sites within the II-nd, IV-th, VI-th, VIII-th co-ordination shells around a 'zero' site), and Ni-atoms will be localized at the face centres of the unit cell. Evidently, this is why the probabilities of jumps of α-atoms (Fe) into the sites $\{R_{II}\}$, $\{R_{IV}\}$, $\{R_{VI}\}$, $\{R_{VIII}\}$ (**Fig. 3***b*) are higher than probabilities of jumps into the sites $\{R_I\}$, $\{R_{III}\}$, $\{R_V\}$, and $\{R_{VIII}\}$.

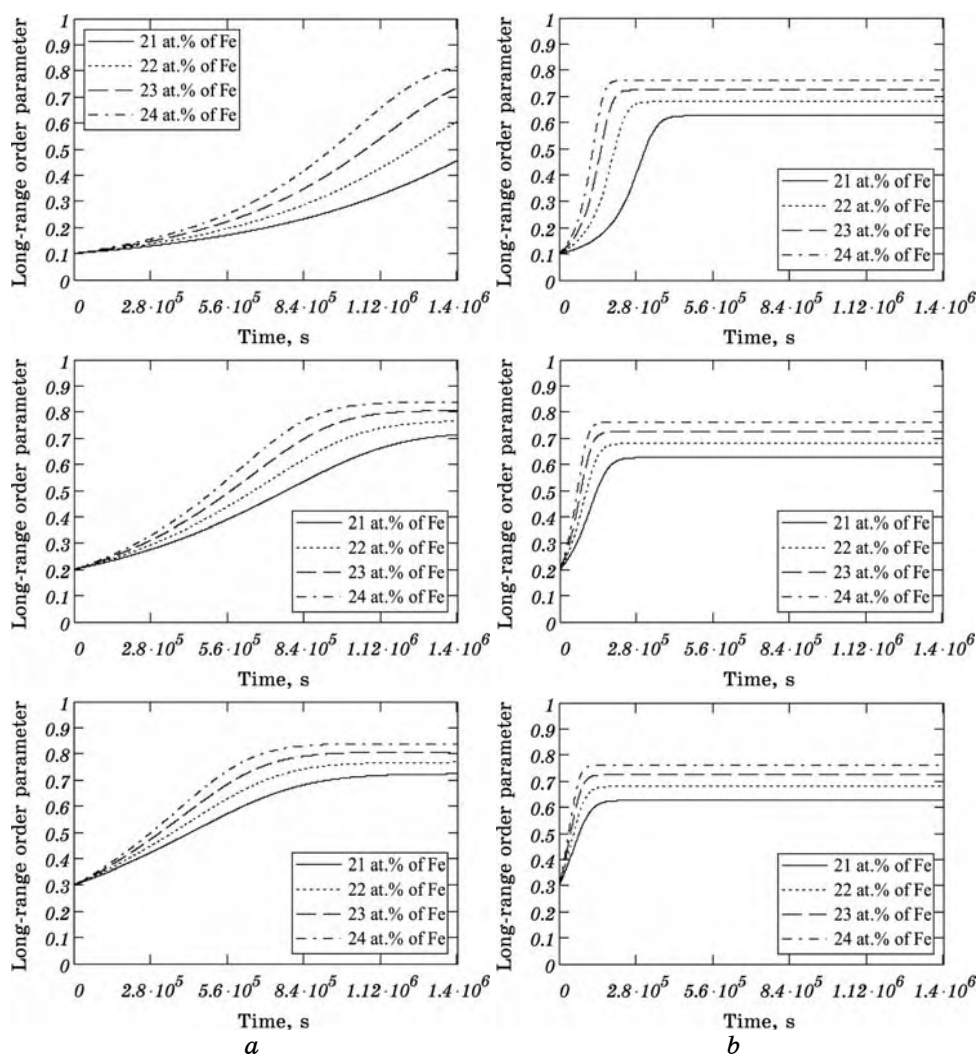

**Fig. 9.** Theoretical dependences of long-range order parameter, η, on the annealing time, $t$, for the non-stoichiometric Ni−Fe Permalloys with different $\eta_0$ at 673 K (*a*) and 743 K (*b*).

The jumps into the sites of the II-nd, IV-th, VI-th, VIII-th co-ordination shells are realised mainly as nearest-distance jumps from the sites where arrangement of α-atoms (Fe) is less energy-wise 'advantageous', *i.e.* from the sites of the I-st, III-rd, V-th, VII-th coordination shells.

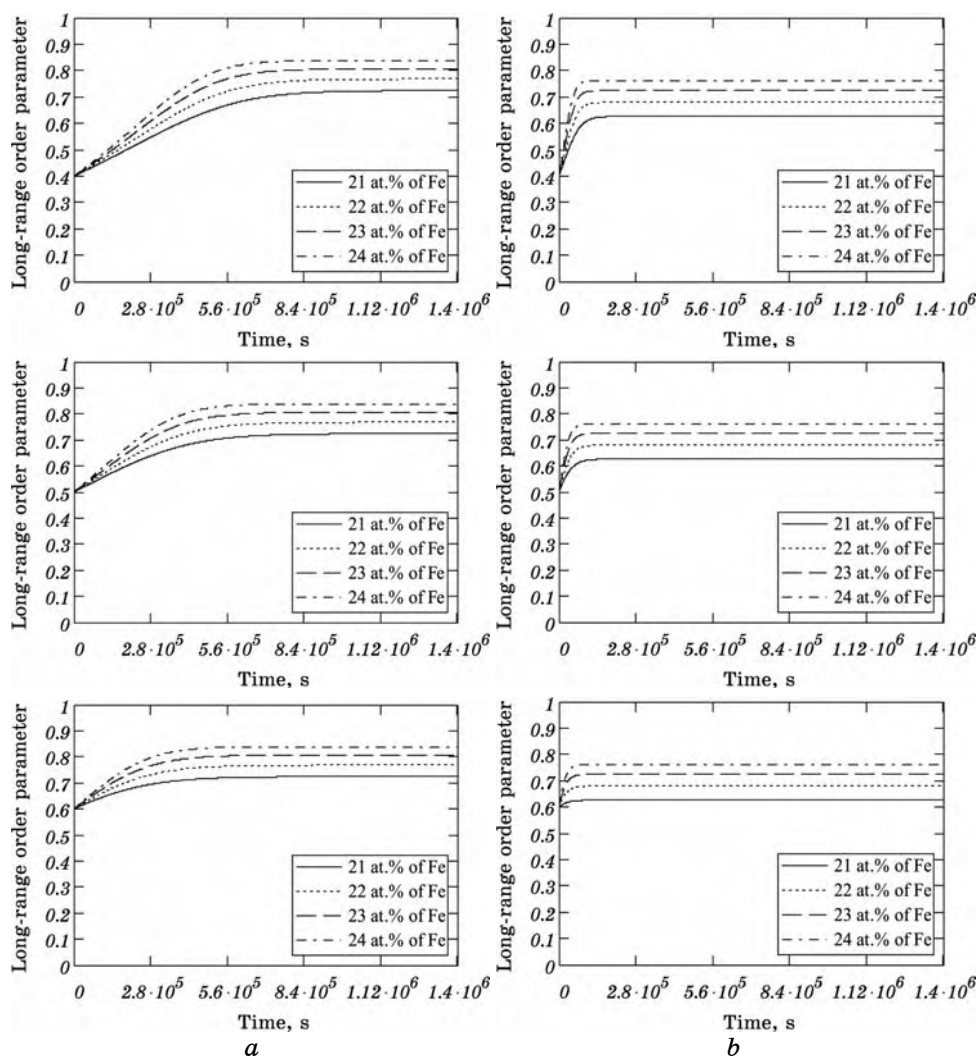

**Fig. 10.** The same as in Fig. 7, but with another $\eta_0$.

Thus, the n-dependence of probabilities presented in Fig. 3*b* is unrepugnant to the vacancy mechanism of diffusion in the alloy at issue.

**iii)** Dependence of the normalized 'potential' function on a radius of co-ordination shell, $R_n$, is non-monotone (Fig. 4). For some $R_n$, the function value is positive, for another one, it is negative. This determines thermodynamic 'disadvantage' or 'advantage' for the 'slow' α-atom to stay in the corresponding sites $\{R_n\}$.

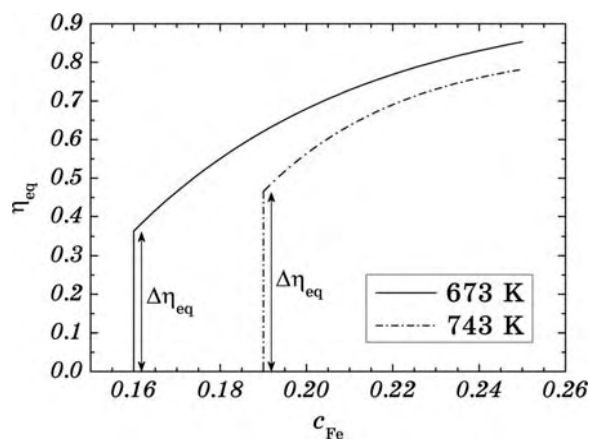

**Fig. 11.** Equilibrium long-range order parameter, $\eta_{eq}$, *vs.* concentration of Fe, $c_{Fe}$, for a Ni–Fe Permalloy at the two annealing temperatures.

The temperature increasing has an ambiguous influence on the 'potential' fields caused by the concentration heterogeneities. 'Potential' field is increased for some co-ordination shells, and it is decreased for other ones (Fig. 4).

As quite clear, if the temperature is elevated, all the probabilities increase as well (Fig. 3). Nevertheless, although the temperature increasing promotes the increasing of any atomic-probabilities jumps generally, but decreasing of the action of 'potential' field generated by the atoms of defined element and caused by its concentration heterogeneities onto the far distant sites (co-ordination shells) results in increasing of the atomic-jumps' probabilities of just this element, first of all, into the sites, which are more distant from the 'source' of heterogeneity (Fig. 4).

**iv)** Total energies of activation of diffusion and self-diffusion of 'slow' (Fe) atoms in disordered $^{62}Ni_{0.765}Fe_{0.235}$ Permalloy are 3.21 eV and 2.56 eV, respectively. Exchange-diffusion migration activation energy of Fe and Ni atoms in long-range ordered $Ni_3Fe$ alloy is proved to be 1.60 eV. The first two values are higher than the third one because the latter does not involve the energy of vacancy formation; migration energy in $Ni_3Fe$ Permalloy is evaluated within the alloy model without vacancies. Within the framework of this model, the vacancy-formation energy is 38% and 50% of the total activation energies of 'slow' (Fe) atoms' diffusion and self-diffusion, respectively.

**v)** Inasmuch as $|L_\odot(R_l)| < |\Lambda_\alpha(R_n)|$, the diffusivities listed in Table 2 are lower than diffusivities presented in Table 1. That is because of some reasons.

Firstly, it is because of a mixing energy dependent on both the tem-perature and the concentration.

Secondly, because the long-range ordering phase is formed at more low temperatures as compared with a disordered phase.

Thirdly, below the order–disorder-transformation temperature, the diffusion mechanism in long-range ordering alloys may be modified, and this will affect the value of $D$ in the direction of the observed variation. In fact, the probability of exchange ('ring') mechanism of diffusion is small. It is proved by the magnitude of Onsager kinetics coefficient.

**vi)** Theoretical curves in Fig. 6*a* are not in a good coincidence with the experimental points [2] in Fig. 5. It is because of two reasons.

Firstly, to calculate and to plot the evolution curves for $\eta = \eta(t)$, we used the time-optimized ('*average*') values (in Table 2) of the Fourier transform of the Onsager kinetics coefficients, $\tilde{L}_\bigcirc(\mathbf{k}_X, T)$, estimated from the experimental values [2] of $\eta$ at the different $T$-annealing times, $t$.

Secondly, all experimental values of $\eta$ (in Fig. 5) are *instantaneous*. They are *not equilibrium* at all (especially at 673 K). Even after annealing during $3.6 \cdot 10^5$ s, experimental value of $\eta$ does *not* reach its *equilibrium* value; that is why it is lower than theoretical one (see Fig. 6*b*).

**vii)** At low temperatures ($T^*$ is lower than $\cong 0.105$), equilibrium long-range order parameter for non-stoichiometric Ni–Fe Permalloys with both $c_{Fe} < 1/4$ and $c_{Fe} > 1/4$ is always lower than for stoichiometric one ($c_{Fe} = 1/4$). However, equilibrium long-range order parameter for non-stoichiometric Permalloys with $c_{Fe} > 1/4$ may be higher than for a stoichiometric one at high temperatures ($T^*$ is higher than $\cong 0.105$) (see Figs. 7, 8).

**viii)** Figures 9 and 10 demonstrate that both the concentration of alloying component (Fe) and the annealing temperature strongly affect the quantitative and qualitative changes of the kinetic and equilibrium paths. The four concentrations and the two temperatures give not only the different profiles of relaxation curves but also the equilibrium long-range order parameter value. The rate of a long-range order parameter changing is decreased and the relaxation time is increased if the concentration of alloying component is decreased (below a stoichiometric composition).

The rate of the long-range order parameter change is higher at the initial stage of annealing. The decreasing of an initial long-range order parameter promotes the increasing of this rate and does not influence nowise on the equilibrium long-range order parameter—its magnitude is the same for all $\eta_0$ at a given $T$ (Figs. 9, 10).

**ix)** *Non-equilibrium* long-range order parameter for $Ni_3Fe$ Permalloy after $3.6 \cdot 10^5$ s of annealing at both 743 K and, especially, 673 K (Fig. 5) confirms once more that any experimental data relating to the structural and physical properties of solid solutions contain directly an information only about their *instantaneous* characteristics. Prediction of their *equilibrium* values is possible only by the theoretical extrapolation (see Figs. 6, 9, 10).

Thus, just an asymptotical estimate of equilibrium values of the physical characteristics from the experimental data about their relaxation kinetics gives a possibility to obtain the quantitatively correct characterization [5–7].

**x)** If atomic fraction of Fe reaches the defined value, the equilibrium long-range order parameter 'jump' at a given temperature of annealing as a Kurnakov's temperature occurs and, below this $c_{Fe}$ value, $\eta$ takes on a zero value (Fig. 11) that is a low-$c_{Fe}$ Permalloy is disordered in the *equilibrium* conditions.

It confirms that order–disorder phase transformation in a Ni–Fe Permalloy is a first-order kind phase transition.

The kinetics and thermodynamical models give the equal equilibrium long-range order parameter values as well as the equal order–disorder transition point.

**xi)** Using the equation of kinetics (4), we can forecast the evolution of morphological imaging of the structural configurations described by the single-site atomic-occupation probabilities. For instance, a computer simulation technique describing these evolution processes was developed in [14–20], and opportunity of growth of intermediate metastable phases as well as formation of antiphase $L1_2$ domains were proved in [21]. In these all works [14–21], the results were obtained only for so-called 'timesteps' of $T$-dependent reduced time, $t^*$, because of an uncertainty of the Onsager-type kinetics coefficients. The last values are evaluated in this paper; therefore, the real time ordering-process analysis is possible now within the actual $t$ scale.

**xii)** The model used here for the characterization of 'macrodiffusion' by means of 'microdiffusion' parameters obtained from the independent data about the short-range or long-range order kinetics relates to the Permalloys with a fixed volume. However, the pressure fixation is more convenient in a practice; most of the processes occur under the fixed pressure (*e.g.*, atmosphere pressure, Earth's interior pressure, *etc.*). Thus, additional investigations are required taking into account an influence of the external pressure on the short-range and long-range ordering processes in alloys.

## ACKNOWLEDGEMENTS

The work was performed within the framework of the project supported by the NATO Reintegration Grant (RIG 981326) and Scholarship of the World Federation of Scientists, which are gratefully acknowledged by the first author.